\documentstyle[12pt,epsf]{article}
\begin{document}
\newcommand{\half}{\mbox{$\frac{1}{2}$}}
\begin{titlepage}

\newcommand\letterhead {%
\hfill\parbox{8cm}{    \large \it UNIVERSITY of PENNSYLVANIA\\
\large Department of Physics\\
David Rittenhouse Laboratory\\
Philadelphia PA 19104--6396}\\[0.5cm]
{\bf PREPRINT UPR--0127MT}\\ July 1993}
\noindent\letterhead\par
\vspace{2 cm}
\noindent
\begin{center}
{\Large \bf The Skyrme model description of
 the spin-orbit force}\\
\end{center}
\begin{center}
{\bf \large
R. D. Amado, Bin Shao and
Niels R. Walet}
\end{center}
\begin{center}
\vfill
{\em Submitted to Physics Letters B}
\end{center}
\end{titlepage}
\title{
The Skyrme model description of
 the spin-orbit force}

\author{
R. D. Amado, Bin Shao and
Niels R. Walet\thanks{address after Sept.~1, 1993:
Instit\"ut f\"ur theoretische Physik III, Universit\"at Erlangen-N\"urnberg,
D-91058 Erlangen, Germany}\\
Department of Physics, University of Pennsylvania,\\
Philadelphia, PA 19104}
\date{\today}
\maketitle

\begin{abstract}
We examine the derivation of the spin-orbit force from the
Skyrme model.  We find
substantial agreement with phenomenological
potentials.
We show that a systematic and simultaneous treatment of all
components of the kinetic energy introduces a minus sign
relative to previous calculations for the iso-spin independent
part of the interaction.
We also explain the smallness of the iso-spin dependent part
of the spin-orbit potential.
\end{abstract}

The challenge of obtaining the $NN$ interaction from a QCD inspired
starting point has recently been given new impetus by results
in the Skyrme model \cite{NN,VinhMau}.
Careful treatment of the non-linear nature of
the model and of quantum corrections leads to significant central attraction,
in substantial agreement with phenomenology.  There is even a strong short
range repulsion, though its detailed description certainly lies
outside the scope of the Skyrme model with pions only.
Combined with the well known
Skyrme treatment of one pion exchange and the tensor force, all this yields
a satisfactory picture of the static part of the $NN$ interaction that
gives a good description of the  phase shifts.
The next step is to consider the non-static parts of the interaction.
Here we consider
the spin-orbit force.  The dominant part of
the phenomenological spin-orbit interaction is iso-spin independent.
Attempts to reproduce it from the Skyrme model have given the correct
order of magnitude, but the wrong sign \cite{RiskaDanbom,Otofuji}.
In this note we point out that a careful treatment of the conversion
from velocities to canonical momenta in fact reverses that sign and
hence yields a spin-orbit interaction in substantial agreement
with the data.  We also show that a small iso-spin dependent
spin-orbit force emerges naturally.

The principal ingredients in the successful
static $NN$ interaction calculations
are numerically ``exact'' solutions for the baryon number two
($B=2$) system at finite separation, and quantum corrections implemented
through
the Born-Oppenheimer approximation. The exact results differ
considerably from earlier ones based on the simple product Ansatz
largely because the static interaction energy arises as the relatively
small difference of two large numbers.  Cancellations must therefore be
treated with care.  Although a full treatment of the non-static terms
should also be done exactly, there are no corresponding cancellations in
this case, and one can hope to get insight from the
product Ansatz.  It is for that reason that the wrong sign for
the spin-orbit interaction obtained in earlier product Ansatz
calculations was particularly disappointing \cite{RiskaDanbom,Otofuji}.

Since there has not been a complete study of the kinetic energy of the
Skyrme model, even though pieces can be found in the literature
\cite{RiskaDanbom,Otofuji,Oka},
we decided to undertake such a study.
A part of our purpose was to understand the puzzle of the spin-orbit sign.
In this note we sketch the steps in our evaluation, and show that
there is a simple solution to the sign problem.

The Skyrme model \cite{Skyrme,Reviewpapers}
is a non-linear field theory that can be realized in
terms of an SU(2)-valued matrix field $U$, with Lagrangian density
\begin{equation}
{\cal L} = \frac{f_\pi^2}{4} {\rm Tr}[
\partial_\mu U(x) \partial^\mu U^\dagger(x)]
+ \frac{1}{32g^2}{\rm Tr}[ U^\dagger\partial_\mu U, U^\dagger\partial_\nu U]^2
+\frac{f_{\pi}^2 m_\pi^2}{4}[{\rm Tr}(U+U^\dagger)-4].
\end{equation}
The model is covariant, as well as invariant under global
SU(2)-rotations that are identified with the iso-spin symmetry.
The only parameters in the model are the pion mass, $m_{\pi}$, the pion
decay constant, $f_{\pi}$, and the dimensionless parameter, $g$.  It is
customary not to take $f_{\pi}$ from experiment but rather to adjust
$f_{\pi}$ and $g$ to give the nucleon mass and the nucleon-delta
splitting correctly.  We take $f_{\pi}= 6.45$ MeV,
$g= 4.48$ and use $m_{\pi}= 138$ MeV.   It should be kept
in mind that all the results on the static $NN$ force and on the
spin-orbit interaction are based on only these three parameters.
As was
discovered by Skyrme the model has a topologically conserved quantum number,
which is identified as the baryon number, $B$.
The $U$ field is interpreted as a combination of a scalar $\sigma$ field
and an iso-vector pion  field, $U=(\sigma +
i\vec{\tau} \cdot \vec{\pi})/f_\pi$.
The $\sigma$ field is not an independent physical field due to the unitarity
constraint on $U$.

The standard time-independent solution to the classical field equations
for $B=1$ is the defensive hedgehog, where the pion field points radially
outward,
\begin{equation}
U_1(\vec{r}) = \exp( i\vec{\tau} \cdot \hat{r} f(r) ).
\end{equation}
The baryon number of this state is given by $B=(f(0)-f(\infty))/\pi=1$.
This solution breaks translational invariance, as well as the $O(4)$
spin-iso-spin symmetry, but the sum of spin and iso-spin
is still conserved.
If we perform a global SU(2) iso-rotation on the state,
\begin{equation}
U_1(\vec{r}|A) = A^\dagger U_1(\vec{r}) A,
\end{equation}
we obtain a state of the same energy.

For the $B=2$ system, we will  use the product Ansatz.
This Ansatz makes use of the fact that the product of two $B=1$ solutions
has baryon number two. The most general Ansatz we can  construct
from  two hedgehogs consists of the product of two separated and
rotated hedgehogs,
\begin{eqnarray}
U_2(r|\vec{R} AB)&=& A^\dagger U_1(\vec{r}-\vec{R}/2) A
                     B^\dagger U_1(\vec{r}+\vec{R}/2) B
\nonumber\\
                 &=& U_2(r|\vec{R} CD)
\nonumber\\
                 &=& D^\dagger C^{1/2\dagger} U_1(\vec{r}-\vec{R}/2) C
                     U_1(\vec{r}+\vec{R}/2) C^{1/2\dagger} D.
\label{eq:PA}
\end{eqnarray}
In the last line of (\ref{eq:PA}) we have introduced the matrix $D$,
that describes the rigid iso-rotation of the whole system,
as well as a relative iso-rotation $C$. When $R$ is very large changing $C$ or
$D$ does not change the energy of the solution. For smaller $R$, $D$
still generates
a zero-mode (corresponding to broken iso-spin symmetry), but the energy will
depend on $C$. The energy is also invariant under spatial rotation,
due to the conservation of angular momentum $\vec{J}=\vec{L}+\vec{S}$.

In order to evaluate the kinetic energy, we make all the collective parameters
($\vec{R},A,B$) time dependent, and substitute the product Ansatz (4)
into the Lagrangian.
This leads to an effective Lagrangian containing a kinetic energy
that is second order in the time
derivatives due to the special nature of the
Skyrme Lagrangian. Note that we have
already used Galilean invariance to decouple the center-of-mass motion from
the relative motion.

We have evaluated all terms in the kinetic energy.
The terms can be expanded
in  tensors in $C$ and powers of $\hat R$
(as well as the velocities, of course).
The tensors form a basis for  the
irreducible representations of the $O(4)$ formed from the four
components of $C$.
We find only symmetric representations of
$O(4)$ appearing, $(\sigma,0)$, with $\sigma$ even.
Only $\sigma =0$ (the unit operator),
$\sigma =2$ and $\sigma =4$ occur in the expansion.
Each of these terms is multiplied by a coefficient function that depends only
on the size of $R$. We find typically that these functions are numerically
largest if $\sigma =0$, smaller
if $\sigma =2$, and smallest if $\sigma =4$. Note that the
$\sigma =4$ operators
have zero matrix element in two nucleon states.
We also find several non-zero coefficients that are not even under
parity or the interchange of the two Skyrmions -- a well known disease
of the product Ansatz. In the remainder of the discussion we
ignore these terms. The full expansion of the kinetic energy is
quite complicated and will be reported on separately \cite{SWA}.
Here we concentrate on the  spin-orbit coupling and only consider
terms that are dominant in that expansion.

In order to quantize the kinetic energy, we must first
invert the mass matrix to obtain the Hamiltonian \cite{Goldstein}.
In other words we perform the standard transformation from a
Lagrangian,
\begin{equation}
L = \half \sum_{ij} \dot q_i M_{ij} \dot q_j -V(\vec q),
\end{equation}
to a Hamiltonian,
\begin{equation}
H = \half \sum_{ij} p_i (M^{-1})_{ij} p_j + V(\vec q),
\end{equation}
with $ p_i = \sum_j M_{ij} \dot q_j$ conjugate to $q_i$.
We separate the mass matrix in terms of its dependence
on the $O(4)$ label, $\sigma$,
\begin{equation}
M = M_0 + M_2  +M_4.
\end{equation}
As mentioned above these terms decrease rapidly in size with
increasing $\sigma$.  We take advantage of this fact to drop
$M_4$ and write
\begin{equation}
M^{(-1)}  \approx  M_0^{-1} - M_0^{-1} M_2 M_0^{-1} + {\rm small~terms}
\end{equation}
The $LS$ coupling already appears in $M_0$, as an off-diagonal terms
coupling the velocity $\vec v = \dot{\vec R}$ to the sum
of the two rotational velocities for the individual Skyrmions.
If we denote the sum and difference of the rotational velocities
 by $\vec\omega^\pm$, we find that the {\em dominant}
contribution of $M_0$ to the kinetic energy is given by
\begin{equation}
K = \half\{2 d_1^v(R) v^2 + d_1^\omega(R) ({\omega^+}^2+{\omega^-}^2)
          + 2 u_3^{v\omega}(R)(\vec v \times \vec R
	 \cdot  \vec \omega^+)\}.
\label{eq:K0}
\end{equation}
Here the quantities $d_1^v,d_1^\omega,$and $u_3^{v \omega}$ are
functions of $R$, calculated from the Skyrmion configurations.  We
have that
$d_1^v \stackrel{R\rightarrow\infty}{\longrightarrow} M/4$,
$d_1^\omega \stackrel{R\rightarrow\infty}{\longrightarrow} 2\Lambda$ and
$u_3^{v\omega} \stackrel{R\rightarrow\infty}{\longrightarrow} 0$
($M$ is the mass of a single Skyrmion, and
$\Lambda$ its moment of inertia.)

The approach in \cite{RiskaDanbom,Otofuji} was to replace
$\vec v$ by $\vec p/(M/2)$, and $\vec \omega^+$ by $\vec S/(2\Lambda)$
to find
\begin{equation}
V_{LS} = \frac{u_3^{v \omega}(R)}{M\Lambda} {\vec L}\cdot \vec S.
\end{equation}
However this procedure is {\em not} equivalent to
inverting the mass matrix. The matrix to be inverted here is basically
two-by-two with the spin-orbit coupling
coming from the off diagonal part.  As is
well know the off diagonal terms in such a case change sign under inversion.
If we carry out the inversion from (5) to (6) we
indeed find for the spin orbit force
\begin{equation}
V_{LS} = -\frac{u_3^{v \omega}}{2d_1^v d_1^\omega-(u_3^{v\omega})^2}
 {\vec L}\cdot \vec S
\stackrel{R\rightarrow\infty}{\longrightarrow}
-\frac{u_3^{v \omega}(R)}{M\Lambda} {\vec L}\cdot \vec S.
\end{equation}
Thus we find the {\em correct} sign for the spin-orbit interaction.
\begin{figure}
\epsfysize=8cm
\centerline{\epsffile{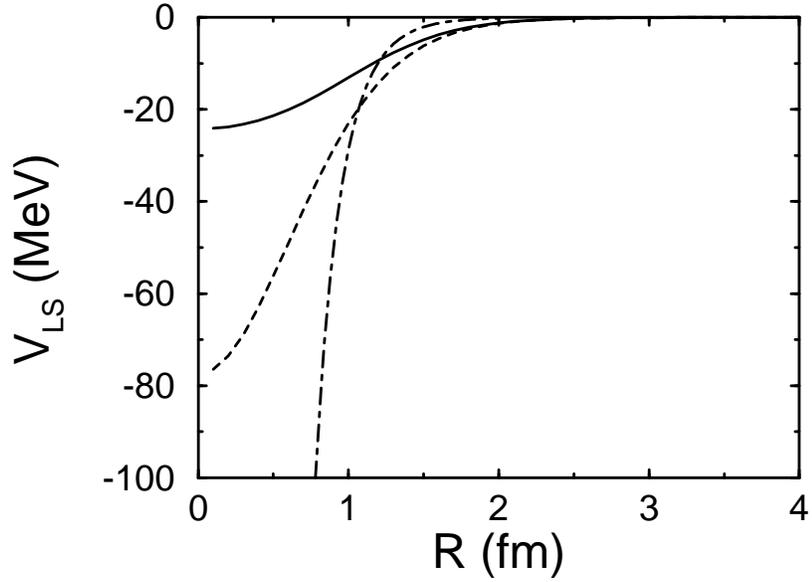}}
\caption{The isospin independent spin-orbit potential as a function
of nucleon separation $R$.
The solid line is calculated using
the first expression in Eq.~(11), whereas the dashed
line represents the result from the second expression. The dash-dotted line
is the corresponding term in the BonnR potential \protect{\cite{Bonn}}.}
\end{figure}
In Fig. 1 we show our calculation for the iso-spin independent part
of the spin-orbit interaction.  We show two
calculated curves compared with the BonnR potential \cite{Bonn}.  One
calculation uses
the asymptotic values (large $R$) for $d_1^v$ and $d_1^{\omega}$ for all
$R$ and neglects the $u_3^{v \omega}$ in the denominator (this is close to
the spirit of \cite{RiskaDanbom}), while the other keeps all the terms in
Eq. (11) with their full $R$ dependence.  Both curves agree fairly well with
phenomenology beyond 1 fm, where the product ansatz makes sense, but differ
at small distances. The $R$ dependent curve is probably the better indicator of
what a full calculation will yield.

\begin{figure}
\epsfysize=8cm
\centerline{\epsffile{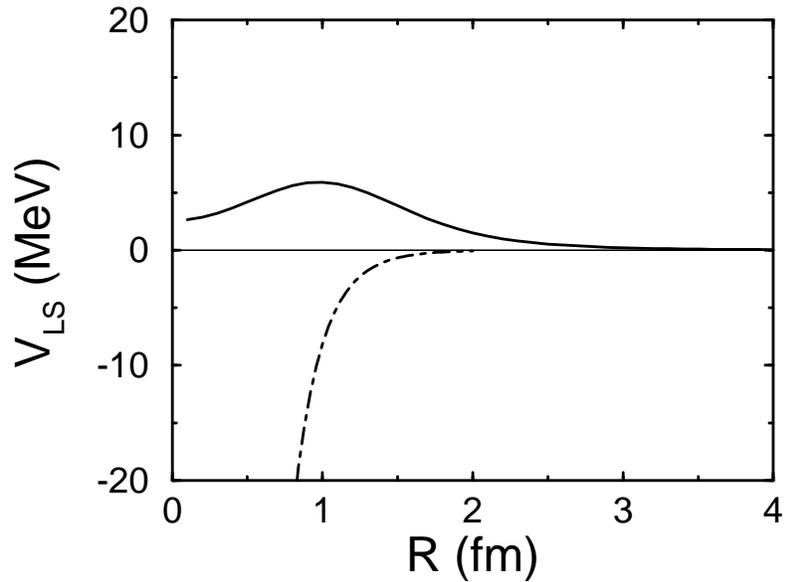}}
\caption{
The isospin dependent spin-orbit potential as a function
of nucleon separation $R$.
The solid line is  our result, and
the dash-dotted line
the corresponding term in the BonnR potential.}
\end{figure}

The iso-spin dependent part of the spin-orbit interaction comes from the
$M_2$ term in Eq. (8). It is shown in Fig. 2 and compared with the
BonnR potential  \cite{Bonn}.
As discussed above it is smaller than the iso-spin
independent part, as is the data. It is also complicated with many terms
contributing.  As seen in the figure, it is positive at large $R$,
and oscillates at smaller $R$.  This oscillation
comes from a cancellation among many terms and is probably quite
sensitive to model assumptions.
This sensitivity to cancellations recalls the static interaction in
the central channel, suggesting that channel coupling
may again be important \cite{NN}.

Riska and Schwesinger \cite{RiskaSchwesinger}
have added a 6th order term to the Skyrme model,
and claim to find the right sign for the
spin-orbit interaction. The results
of this note shows that their discussion is faulty.
It is not necessary to include a 6th order term in the
Skyrme model to find agreement between the model and phenomenology
for spin-orbit coupling. Furthermore their
6th order term calculated as in (6) and (7),
will lead to the {\em wrong} sign of the
spin orbit interaction.

In summary we have shown that the Skyrme model yields the correct sign
and order of magnitude for the $NN$ spin-orbit interaction.  Coupled with
our earlier demonstration that the Skyrme model correctly accounts for the
main features of the static $NN$ interaction, one pion exchange,
mid-range attraction, tensor force, short range repulsion, the result
concludes successfully the first steps in the derivation of the
$NN$ interaction from a starting point based in non-perturbative
QCD. \\[.25in]

This research was supported in part by the U.S. National Science Foundation.

\newpage

\end{document}